# Vowel-based Meeteilon dialect identification using a Random Forest classifier

Thangjam Clarinda Devi
*Department of Computer Science and Engineering*
*Indian Institute of Information Technology Manipur*
Manipur, India
thangc@iiitmanipur.ac.in

Kabita Thaoroijam
*Department of Computer Science and Engineering*
*Indian Institute of Information Technology Manipur*
Manipur, India
kabita@iiitmanipur.ac.in

*Abstract*—This paper presents a vowel-based dialect identification system for Meeteilon. For this work, a vowel dataset is created by using Meeteilon Speech Corpora available at Linguistic Data Consortium for Indian Languages (LDC-IL). Spectral features such as formant frequencies (F1, F1 and F3) and prosodic features such as pitch (F0), energy, intensity and segment duration values are extracted from monophthong vowel sounds. Random forest classifier, a decision tree-based ensemble algorithm is used for classification of three major dialects of Meeteilon namely, Imphal, Kakching and Sekmai. Model has shown an average dialect identification performance in terms of accuracy of around 61.57%. The role of spectral and prosodic features are found to be significant in Meeteilon dialect classification.

## I. Introduction

A dialect is a certain form of a language which is unique to a particular region or social group [1]. The study of dialectal variations effect on the performance of an Automatic Speech Recognition (ASR) is a major research topic. The accuracy of an ASR can be improved to a large extent by including the knowledge about the dialects in pronunciation and acoustic model training [2].

Dialect specific information exists at both the segmental and supra-segmental level of a speech signal. Majority of the existing automatic dialect identification systems have been proposed by using both segmental and supra-segmental acoustic features. Phonotactic approach has also been considered in the literature. Acoustic features such as Mel frequency cepstral coefficients (MFCC), Linear predictive coding (LPC), Shifted Delta coefficients (SDC) and Formant frequencies have been studied in the past [3]. Similarly, prosodic features are modelled from the pitch, energy and duration values [3].

Majority of the existing work has been done for high resourced languages such as English, Chinese, Japanese, Dutch, Arabic, and so on. But many Indian regional languages are low-resourced except for Hindi language. Due to this reason, research on dialect identification for such languages are still in its developing stage. Very few previous attempts [4] have been made in building Meeteilon dialect identification systems. Meeteilon is the official language of the state of Manipur. It was included in the Eighth Schedule to the Constitution of India on 20th August, 1992 [5]. Meeteilon has dialects of Imphal, Kakching, Sekmai, Andro, Khurkhul, Phayeng, Kwatha, Thanga, etc. Imphal dialect is considered the standard form of speech [6]. The present study focuses only on three dialects namely, Imphal, Kakching and Sekmai. The Imphal dialect, spoken around the capital city, is considered to be the standard variety. The speakers of Kakching dialect live in Kakching, a district which is situated to the south of Imphal valley at a distance of about 45 kms. The speakers of Sekmai dialect live in Awang Sekmai, situated near Imphal-Dimapur road about 19 kms from Imphal in the North. According to phoneme inventory, Manipuri has 38 phonemes- 36 segmental and 2 suprasegmental phonemes. Out of these 38 phonemes, 24 are consonants, 6 are monophthongs, 6 are diphthongs and the remaining 2 are tones. It is agreed that the language has 15 (/p/, /pʰ/, /t/, /tʰ/, /c/, /k/, /kʰ/, /m/, /n/, /ŋ/, /s/, /h/, /l/, /w/ and /j/) indigenous consonants and the remaining 9 (/b/, /bʰ/, /d/, /dʰ/, /g/, /gʰ/, /ɟ/, /ɟʰ/ and /ɹ/) are exotic consonants. The six monophthongs are /i/, /e/, /ə/, /a/, /u/ and /o/ and the six diphthongs are /əi/, /əu/, /ai/, /au/, /oi/ and /ui/ [7].

Rest of the paper is organized with the following sections: The description of the Meeteilon vowel dialect dataset is discussed in Section 2. Vowel based dialect identification system is described in Section 3. Section 4 provides the experimental results. Finally, Section 5 contains conclusion and scope for future works.

## II. Dialect Dataset

In this study, Manipuri Speech Corpora collected from Linguistic Data Consortium for Indian Languages (LDC-IL) [8] [9] is used. The sentences consist of words where the phonemes can occur in various positions and are meant to show the dialectal variants of the speakers. The dataset contains speech files from 150 speakers and each speaker has uttered 25 randomly selected sentences out of 208 unique sentences. Age of the speakers varies from 16 to 51 years. Manual annotation at phoneme level is done to identify the steady-state vowel regions of all the vowels present in the dataset using Praat speech analyzer toolkit [10]. Fig.1. shows an example of an annotation.

As the process of annotation is quite time-consuming, annotation at phoneme level has been done only for the subset of the speech corpora. Thus, for the preparation of the vowel dialect dataset, only a subset of the speech corpora is used. Brief details of dataset used in this study is shown in Table 1. A total of 1080 vowels have been analysed from the sentences uttered by forty five speakers. Almost equal distribution of gender in each dialect has been taken into consideration. Only monophthong vowels such as /ə/, /e/, /i/, /o/, /u/ and /a/ are considered in this study.

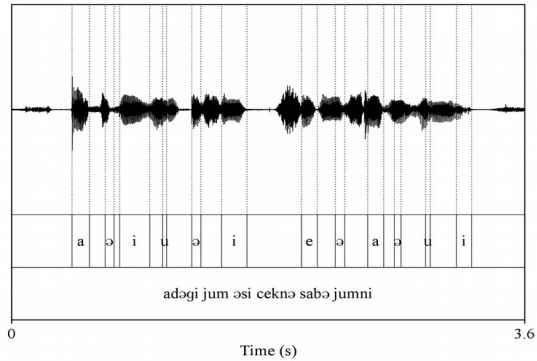

Fig. 1. 2-tier manual annotation of the speech data using Praat.

TABLE I. VOWEL DATASET DETAILS

| Sl. no. | Dialect | Total no. of vowels | No. of speakers (M+F)[a] |
|---|---|---|---|
| 1. | Imphal | 320 | 7+8 |
| 2. | Kakching | 370 | 8+7 |
| 3. | Sekmai | 390 | 7+8 |

[a.] Table shows the description of the vowel dataset (M:Male, F:Female)

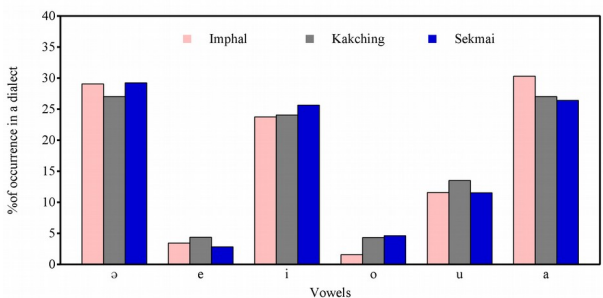

Fig. 2. Percentage of occurrence of each vowel out of a total of 320, 370 and 390 vowels in Imphal, Kakching and Sekmai, respectively.

The frequency distribution of the vowels belonging to the three dialects are shown below in Fig. 2. There are 6 vowels i.e., /ə/, /e/, /i/, /o/, /u/ and /a/ occurring in all the three dialects.

III. VOWEL BASED DIALECT IDENTIFICATION SYSTEM

Spectral and prosodic features are extracted from vowel sounds and used to develop dialect identification system to classify three Meeteilon dialects. Fig. 3 shows an overview of the process in the development of the system. Instead of single classifier based method, ensemble method with decision trees as the base classifiers has been used in this study for the classification of three dialects. Ensemble algorithms are expected to improve the predictive performance of classifier since performances are relying on decisions made by multiple algorithms [11]. Details of feature extraction and implementation of Random forest classifier are given in this section.

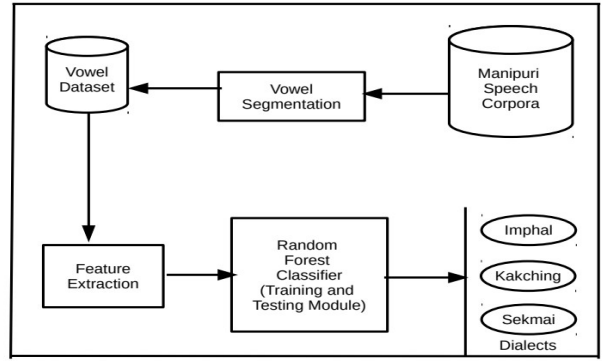

Fig. 3. Vowel based dialect identification system.

A. Feature extraction

The feature extraction is done by using Praat toolkit. For spectral features, formant frequencies (F1, F2, and F3) are extracted from vowel steady state regions. While F1 gives information about vowel height, F2 provides information on vowel front or back position and F3 is commonly used for discrimination of rounded and un-rounded vowels [11]. Fig.4 shows a traditional F2-F1 (in Hertz) plot, drawn using mean values of F2 and F1 of all six vowels from 3 dialects.

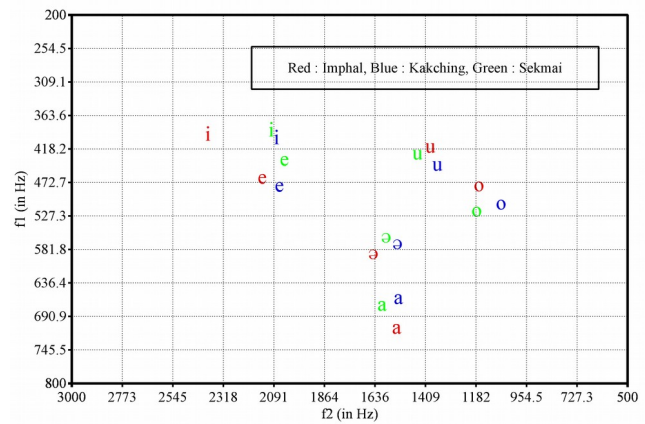

Fig. 4. Average vowel plots of 3 dialects with F2 and F1.

For the prosodic features, segment duration, energy and pitch (F0) values are extracted from the vowel segments of the speech signal. The duration is the length of the vowel segment which corresponds to the speaking rate of the speakers and it is measured in milliseconds. Energy values corresponds to the loudness perceptual property and it plays a prominent role in human auditory perception. Pitch is an important parameter to measure dialectal characteristics and the sequence of fundamental frequency (F0) values constitute pitch contour. The feature extraction process and its analysis have been presented elaborately in our previous work [12].

Six values of each feature (F1, F2, F3, pitch, and energy) are considered. Later, the duration value, intensity value and gender information is added to the vector. In totality, thirty three features are used for representing the input feature vector.

*B. Random Forest Classifier and its implementation*

A Random Forest classifier is one of the most powerful machine learning algorithm. It can train a group of Decision Tree classifiers, each on a different random subset of the training set. To make predictions, it just obtains the predictions of all individual trees, then predict the class that gets the most votes. It uses bagging and feature randomness when building each individual tree to try to create an uncorrelated forest of trees whose prediction by committee is more accurate than that of any individual tree [13]. Bagging (bootstrap aggregation) algorithms combine predictions from independent base models derived from bootstrap samples by sub-sampling through replacement of the original data. The pre-requisites for random forest to perform well are:

• There needs to be some actual signal in our features so that models built using those features do better than random guessing.

• The predictions (and therefore the errors) made by the individual trees need to have low correlations with each other.

While the algorithm itself via feature randomness tries to engineer these low correlations for us, the features we select and the hyper-parameters we choose will impact the ultimate correlations as well. Fig 5. shows how training of a Random Tree classifiers is done.

After feature extraction, the features were then prepared for the classifier by handling text and categorical attributes, the later were One-Hot-Encoded. Then the feature set was divided into training set and test set by stratified sampling in the 80:20 ratio. The splitting is random, using always the same seed. The stratified sampling divides the data into homogeneous subsets called strata, and the right number of instances is sampled from each stratum to guarantee that the test set is representative of the overall data [14]. Random forest classifier contains built-in feature selection, i.e., the model will only include predictors that help maximize accuracy.

In this work, the implementation of the Decision tree based Random Forest classifier is done using RandomForestClassifier class from Python framework (Scikit-Learn) [15]. Instead of tuning the hyper-parameters manually, a good combination of hyper-parameter values are found by using Scikit-Learn's Grid Search. Hyper-parameter values such as max features and n estimators are assigned 12 and 400 respectively. Splitting of a tree node is controlled through the best split decided by Gini criterion among the random subset of features. After the construction of entire decision trees forest, the classifier predicts the dialect by taking the majority votes from the predictions of different trees. Another great quality of Random Tree classifier is that it makes it easy to measure the relative importance of each ——feature. It does that by looking at how much the tree nodes that use that feature reduce impurity on average (across all trees in the forest). It is a weighted average, where each node's weight is equal to the number of training samples that are associated with it. Scikit-Learn computes this score automatically for each feature after training to determine the final importance of the feature, then it scales the results so that the sum of all feature importance is equal to 1. The irrelevant features can then be removed, basically can help in performing feature scaling and then model can be re-trained. The features that are selected at the top of the trees are in general more important than features that are selected at the end nodes of the trees, as generally the top splits lead to bigger information gains.

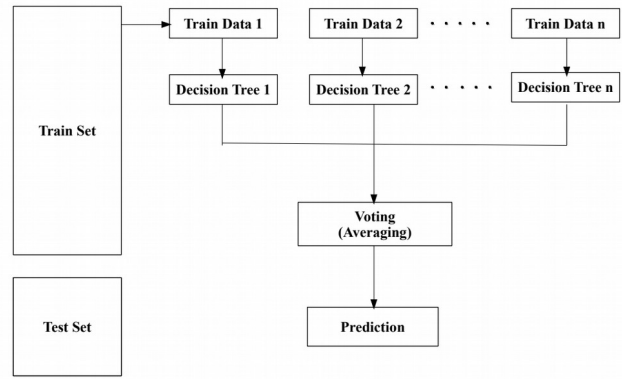

Fig. 5. Training of a Random Tree classifier

IV. EXPERIMENTAL RESULTS

To study the influence of spectral and prosodic features on dialect identification system, the evaluation of the performance of the classifier has been done in three phases. In the first phase, the normalized confusion matrix is obtained after the training is done by using only spectral features i.e., Formant frequencies (F1, F2 and F3) giving an average accuracy of 43.05% shown in Fig 6. The diagonal elements represent the number of points for which the predicted label is equal to the true label, while off-diagonal elements are those that are mislabeled by the classifier. The higher the diagonal values of the confusion matrix the better, indicating many correct predictions. This kind of normalization is useful in case of class imbalance to have a more visual interpretation of which class is being misclassified.

In the second phase, only prosodic features such as F0 values, energy, intensity and duration have been considered. The average performance of the system is observed to be 55.09%. Fig 7. shows the normalized confusion matrix obtained with only prosodic features.

Lastly, third phase measures the performance of the system which uses both spectral and prosodic features. The system is trained with all thirty three features. Fig 8. shows the normalized confusion matrix obtained with both spectral and prosodic features. Results show that average performance of the system is improved and the average accuracy is observed to be around 61.57%. Thus, it indicates that even though vowels have distinct formant frequencies, these are not sufficient to classify the dialects of Meeteilon. Infact, prosodic features have shown higher relevance to dialect identification.

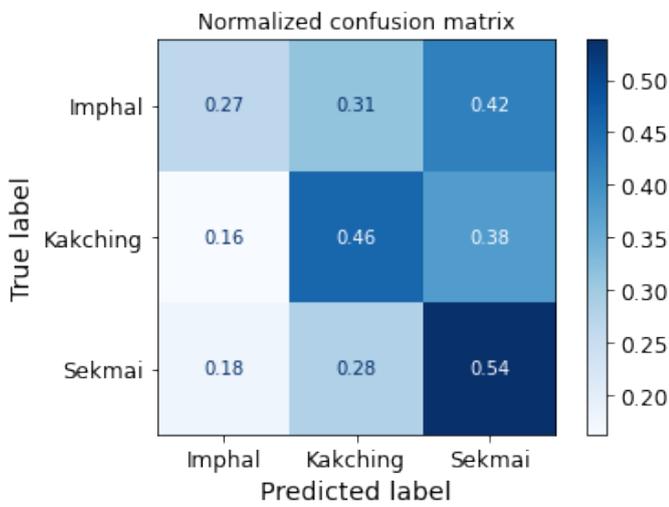

Fig. 6. Normalized confusion matrix obtained with only spectral features for dialect classification.

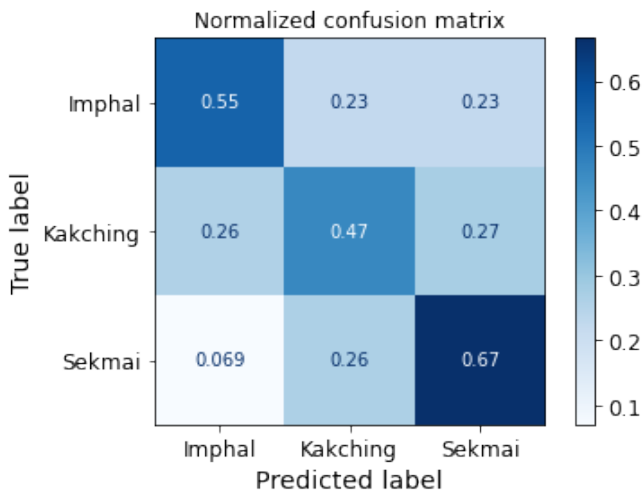

Fig. 7. Normalized confusion matrix obtained with only prosodic features for dialect classification.

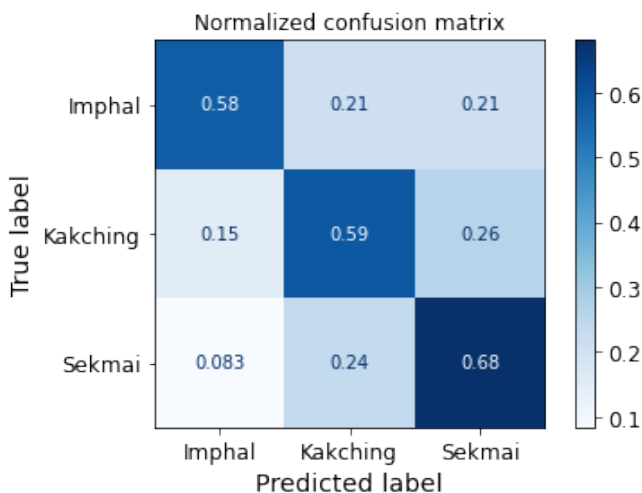

Fig. 8. Normalized confusion matrix obtained with both spectral and prosodic features for dialect classification.

From the confusion matrices, it is noticed that Sekmai dialect is somewhat correctly classified (68%) with high performance among all others. Vowels of Kakching is classified with only 59% accuracy. Vowels of Imphal dialect is classified with only 58%. In all the three phases, the classifier overfits the training data, thus there is a need to increase the training data. Also, appropriate selection of hyper-parameters of the classifiers may also enhance the performance of the system.

## V. CONCLUSION

This paper has reported the vowel based Meeteilon dialect identification system using a Random tree classifier that classifies three Meeteilon dialects. The creation of vowel dialect dataset from the Meeteilon speech corpora is done by annotating using Praat. Three major dialects namely, Imphal, Kakching and Sekmai have been considered for the study. Spectral (F1, F2 and F3) and prosodic features (F0, duration, energy and intensity) extracted from the input speech were analysed. A Decision tree-based Random tree classifier was implemented to capture dialect specific information from the input features. Average performance of the system with only spectral featuresis found to be 43.05%. Average performance of the system with only prosodic features is found to be 55.09%. The performance of the classifier is enhanced by using both the spectral and prosodic features with an accuracy of about 61.57%. Extensive work with larger dataset is currently under process which would probably show better results. Study of tonal variations across the different Meeteilon dialects by annotating tones in the input would help for in-depth analysis. Some boosting methods like Adaptive Boosting and Gradient Boosting can be explored and it may further improve the system performance.

### ACKNOWLEDGMENT

The authors would like to express sincere gratitude towards TEQIP-III, Indian Institute of Information Technology Senapati, Manipur for providing the financial assistance for this study. Also, our sincere and heartfelt thanks to the whole team of LDC-IL (Linguistic Data Consortium for Indian Languages), Central Institute of Indian Languages, Mysore for creating the Manipuri speech corpora and allowing us to use it for this analysis.